\documentclass{aa}  

\usepackage{upgreek}
\usepackage{txfonts}
\usepackage{graphicx}
\usepackage{upgreek}
\usepackage{xcolor}
\newcommand{\ft}[1]{\mathcal{F}\left\{#1\right\}}
\newcommand{\ift}[1]{\mathcal{F}^{-1}\left[#1\right]}
\DeclareMathOperator*{\argmin}{arg\,min}
\newcommand{\edited}[1]{\textcolor{black}{#1}}
\newcommand{\editedn}[1]{{\textcolor{black}{#1}}}

\begin{document}

   \titlerunning{The SM-SCC: increasing throughput for focal-plane wavefront sensing.}

        \title{The spectrally modulated self-coherent camera (SM-SCC): Increasing throughput for focal-plane wavefront sensing}
        
        \author{S. Y. Haffert\inst{1}\fnmsep\thanks{NASA Hubble Fellow}}
        
        \institute{University of Arizona, Steward Observatory, Tucson, Arizona, United States\\ \email{shaffert@arizona.edu}}
        
        \date{Received April ??, 2021; accepted October ??, 2021}

 
  \abstract
  {The detection and characterization of Earth-like exoplanets is one of the major science drivers for the next generation of telescopes. Direct imaging of the planets will play a major role in observations. Current direct imaging instruments are limited by evolving non-common path aberrations (NCPAs). The NCPAs have to be compensated for by using the science focal-plane image. A promising sensor is the self-coherent camera (SCC). An SCC adds a pinhole to the Lyot stop in the coronagraph to introduce a probe electric field. The pinhole has to be separated by at least 1.5 times the pupil size to separate the NCPA speckles from the probe electric field. However, such a distance lets through very little light, which makes it difficult to use an SCC at high speed or on faint targets. }
  {A spectrally modulated self-coherent camera (SM-SCC) is proposed as a solution to the throughput problem. The SM-SCC uses a pinhole with a spectral filter and a dichroic beam splitter, which creates images with and without the probe electric field. This allows the pinhole to be placed closer to the pupil edge and increases the throughput. Combining the SM-SCC with an integral field unit (IFU) can be used to apply more complex modulation patterns to the pinhole and the Lyot stop. A modulation scheme with at least three spectral channels can be used to change the pinhole to an arbitrary aperture with higher throughput. This adds an additional degree of freedom in the design of the SM-SCC.}
  {The performance of the SM-SCC is investigated analytically and through numerical simulations.}
  {Numerical simulations show that the SM-SCC increases the pinhole throughput by a factor of 32, which increases the wavefront sensor sensitivity by a factor of 5.7. The reconstruction quality of the sensor is tested by varying the central wavelength of the spectral channels. A smaller separation between the wavelength channels leads to better results. The SM-SCC reaches a contrast of $1\cdot10^{-9}$ for bright targets in closed-loop control with the presence of photon noise, phase errors, and amplitude errors. The contrast floor on fainter targets is photon-noise-limited and reaches $1\cdot10^{-7}$. The SM-SCC with an IFU can handle randomly generated reference field apertures. For bright targets, the SM-SCC-IFU reaches a contrast of $3\cdot10^{-9}$ in closed-loop control with photon noise, amplitude errors, and phase errors.}
  {The SM-SCC is a promising focal-plane wavefront sensor for systems that use multiband observations, either through integral field spectroscopy or dual-band imaging.}
  

   \maketitle
%
\section{Introduction}
One of the major goals for the current and next generation of large and extremely large telescopes is the detection and characterization of exoplanets. The giant segmented mirror telescopes in particular have the unprecedented angular resolution and sensitivity needed to directly image Earth-like planets around other stars. With these extremely large telescopes, we may even be able to detect biosignatures in their atmospheres \citep{kiang2018exoplanet}. However, ground-based telescopes face a challenge: Imaging through Earth's atmosphere degrades the spatial resolution due to turbulence. High-contrast imaging (HCI) instruments are designed to overcome these challenges by using extreme adaptive optics (AO) to compensate for atmospheric disturbances and recover the angular resolution \citep{guyon2018exao}. And most, if not all, modern HCI instruments use advanced coronagraphs to remove the influence of starlight.

With the current generation of HCI instruments \citep{macintosh2014gpi, jovanovic2015scexao, males2018magaox, beuzit2019sphere}, we can routinely reach post-processed contrast levels of $10^{-4}$ to $10^{-6}$, depending on the angular distance from the host star. With these contrast levels, we are sensitive to hot and massive self-luminous planets \citep{marley2007hotjupiters}. Even though we are sensitive to massive Jupiter-like planets, only a handful of planets have been imaged and spectroscopically characterized. The results from large surveys that targeted \edited{giant gas planets} on wide orbits indicate that the planet occurrence rate drops sharply from 1 to 10 AU \citep{bowler2015gpoccurence, nielsen2019gpies, fernandes2019occurence}. More planets could be directly imaged if the sensitivity close to the star is improved. 

Non-common path aberrations (NCPAs) are wavefront errors that are not seen by the AO system because they are not in the common optical path. The NCPAs are generated inside the instrument and evolve slowly in time due to environmental changes (e.g., temperature variations) or changes in the gravity vector. The NCPAs create a speckle halo that can mimic exoplanet signals, and, due to their slow temporal evolution, they do not average out over typical observing timescales \citep{martinez2013speckle}. Advanced image processing techniques are necessary to remove the stellar speckles and recover the planet signal. However, post-processing algorithms that depend on spatial diversity, such as angular differential imaging \citep[ADI;][]{marois2006adi}, are not able to remove the speckles efficiently at small angles due to the limited spatial diversity. 

Active compensation of the NCPAs during the observations would sidestep this issue. To achieve active compensation, we need to be able to sense the wavefront at the science focal plane, which is typically achieved with a focal-plane wavefront sensor (FPWFS). Sensing the wavefront at any other plane would still allow for the existence of NCPAs. The self-coherent camera (SCC) is an integrated coronagraph and FPWFS that uses a spatial modulation of the stellar speckles to estimate the complex electric field \citep{baudoz2006scc}. The spatial modulation is achieved by adding an off-axis reference hole in the Lyot stop (LS) of the coronagraph. Only a source that hits the coronagraph's focal-plane mask on-axis will scatter light outside the geometric pupil and therefore into the reference hole. This creates Fizeau fringes for the on-axis source only, which can be used to measure the incoming wavefront in both phase and amplitude. However, the reference hole has to be placed quite far from the pupil edge to uniquely determine the wavefront \citep{galicher2010scc}.

The SCC was originally developed and optimized for space-like conditions \citep{baudoz2006scc,galicher2010scc} and has recently been tested for ground-based conditions \citep{galicher2019sccpalomar,singh2019sccaohalo}. The delay in using an SCC for ground-based telescopes is mainly due to the strength of the intensity modulation, which is quite weak for a classic SCC. In space-based conditions, the coronagraphic dark hole can reach contrast levels of $10^{-8}$ to $10^{-10}$, for which an SCC is well suited. Ground-based telescopes, on the other hand, have a post-coronagraphic contrast of $10^{-3}$ to $10^{-6}$. This prompted the development of several modifications to increase the modulation strength.

The fast atmospheric SCC technique (FAST) modifies the coronagraphic focal-plane mask to increase the throughput at the off-axis pinhole \citep{gerard2018fast}. FAST can run with a much shorter exposure time or, equivalently, work on much fainter objects due to the increased amount of light in the reference pinhole. Another way to increase the strength of the modulation is with the fast-modulated SCC \citep[FMSCC;][]{martinez2019fmscc}, where the reference hole is modulated at high speed (1kHz) to freeze atmospheric speckles. The FMSCC takes two images in succession, one without the modulation and one with the modulation. These two images can be processed to reconstruct the interference term with the reference hole placed much closer to the edge of the pupil, where more light will enter the reference hole. The downside of the FMSCC is that the images will be taken one after the other, which will mean that any time-evolving aberration will not be captured correctly.

In this work I propose using a spectrally modulated SCC (SM-SCC) to increase the modulation strength and to enable SCCs to be used in ground-based telescopes. Instead of modulating the reference hole in time, as done by the FMSCC, the reference hole will be \edited{modulated} along the wavelength dimension. This is ideally suited to applications where dual-band imaging (DBI) is already performed, such as H$\alpha$ DBI \citep{close2014discovery} or methane DBI \citep{rosenthal1996efficient}. I also extend the SM-SCC to multiple wavelength channels with a continuously changing modulation amplitude of the pinhole and the LS. This extension can be used in combination with any integral field spectrograph (IFS).

Section 2 describes a classic SCC, and in Sect. 3 the extension to arbitrary spectral modulation of the SCC is described. Section 4 shows simulations to validate the principle of the SM-SCC. 

\section{The self-coherent camera}

\begin{figure*}
        \centering
        \includegraphics[width=\textwidth]{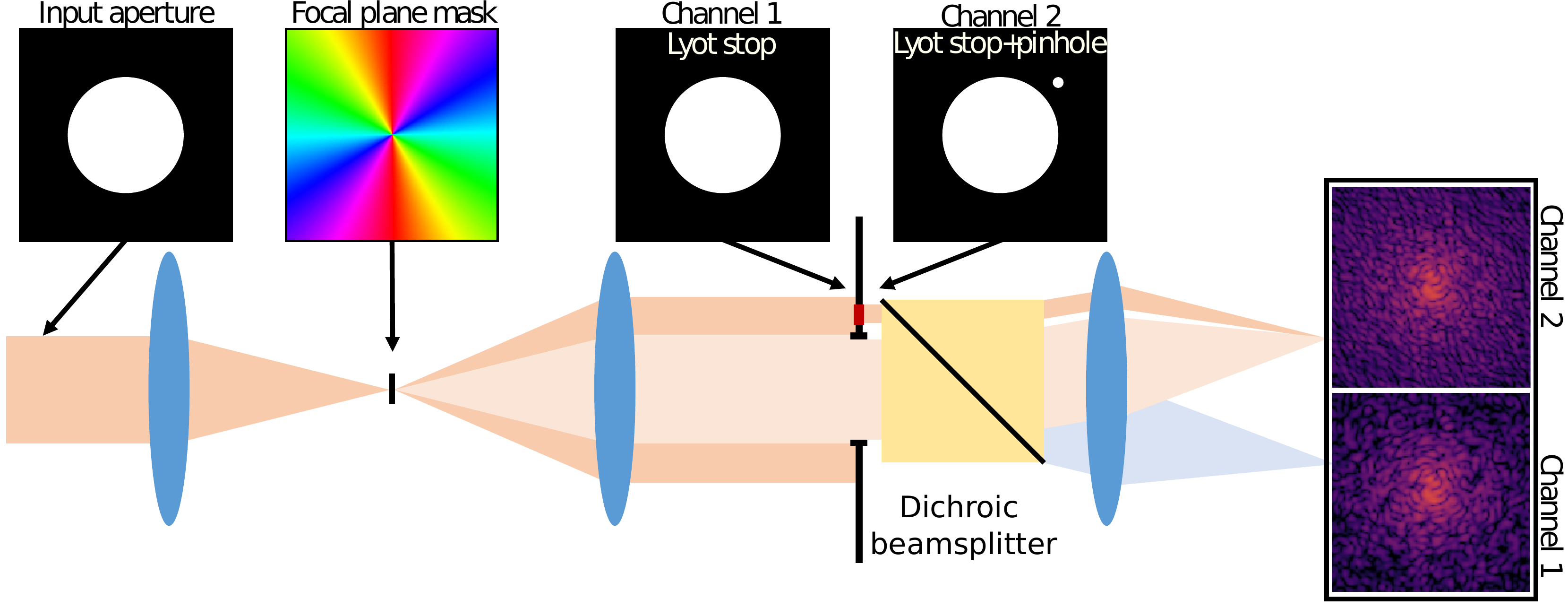}
        \caption{Layout of the (SM-)SCC. The beam is focused on a focal-plane mask that will diffract on-axis light outside of the geometric pupil. Subsequently, a dichroic beam splitter creates two channels: one channel with Fizeau fringes in the focal plane and one channel without. The difference between these two channels will reveal the fringes, which can then be used to retrieve the electric field. For a classic SCC, there is only a single fringed channel.}
        \label{fig:smscc}
\end{figure*}

A standard layout of an HCI instrument with an SCC can be seen in Fig. \ref{fig:smscc}. The incoming stellar light is focused on a focal-plane mask, which scatters light outside the aperture in the LS plane. Any NCPAs will pass through the coronagraph and will leave light within the LS. Similarly, most of the planet light passes through due to the angular offset of the planet, and its light is contained within the aperture in the LS plane. This difference between the planet light and the starlight allows the SCC to modulate the stellar speckles using only the light outside the LS. In the classic SCC setup, a pinhole, with diameter $d_p$, \edited{is placed outside the LS at a distance of $\epsilon$} to create a sinusoidal modulation of the stellar speckles in the focal plane.

The monochromatic focal-plane intensity at wavelength $\lambda$ that is created by an SCC can be described as
\begin{equation}
I = |A_s|^2 + I_{\mathrm{inc}} + |A_r|^2 + 2 \Re\left\{A_s A_r^\dagger \exp{\left(\frac{2\pi i \vec{u}\cdot\vec{\epsilon}}{\lambda}\right)}\right\}.
\label{eq:scc}
\end{equation}
Here, $A_s$ is the focal-plane electric field from the star, $I_{\mathrm{inc}}$ the incoherent contribution from an exoplanet or circumstellar material, $A_r$ the reference electric field from the pinhole, and $\vec{u}$ the position in the focal plane. The $\dagger$ indicates the complex conjugate. The interference term can be extracted from the Fourier transform (FT) of the observed focal-plane image because the modulation frequency is well defined. Taking the FT of the focal plane creates the complex optical transfer function (OTF). The OTF has a diameter twice that of the pupil aperture (or LS aperture) because it is the auto-correlation of the pupil function. The interference fringe creates a sideband, which is the correlation of the pinhole with the pupil. \edited{The sideband has a size of $D + d_p$ at a separation of $\epsilon$.} An example of an OTF can be seen in Fig. \ref{fig:otf_extraction}.

\subsection{SCC requirements}
For a classic SCC, the reference hole requires a minimum distance $(\epsilon_0)$ from the center of the pupil, which was derived in \citet{galicher2010scc}. This distance ensures that the sidebands do not overlap with the central peak of the OTF. The minimum distance is
\begin{equation}
\epsilon_0 = \frac{1}{2}\left(3 + \gamma\right)D.
\label{eq:pinhole_distance}
\end{equation}
Here, $\gamma$ is the relative size of the pinhole with respect to the pupil diameter, $\gamma = d_p / D$. The definition of $\gamma$ that is used throughout this manuscript is the inverse of the definition in previous work. Little light enters the pinhole because the pinhole has to be placed far from the LS edge. The amplitude of the fringes is therefore quite small. This becomes even more of a problem when a larger focal-plane region (more actuators) has to be controlled because a larger region requires a smaller pinhole \citep{mazoyer2013labverification},
\begin{equation}
d_p \leq 1.22\sqrt{2}\frac{D}{N_{\mathrm{act}}}.
\label{eq:pinhole_diameter}
\end{equation}
Here, $N_{\mathrm{act}}$ is the number of actuators across the pupil diameter. The pinhole area \edited{has an inverse dependence on $N_{act}^2$ because the pinhole diameter has to shrink for systems with more actuators. This also means that when more modes have to be sensed because the number of actuators is increased, the amount of light for the reference field will drop.}

\subsection{Wavefront control}
\label{sec:wfs}
Equation \ref{eq:scc} shows that there is a linear relation between the stellar electric field and the interference term. This can be used to estimate the phase and amplitude of coronagraphic residuals. The FT of Eq. \ref{eq:scc} is
\begin{equation}
\begin{split}
\ift{I} &= \ift{|A_s|^2 + I_{\mathrm{inc}} + |A_r|^2} \\
&+ \ift{A_s^\dagger A_r}\otimes \delta(r - \epsilon) \\
&+ \ift{A_s A_r^\dagger}\otimes \delta(r + \epsilon).
\end{split}
\label{eq:otf}
\end{equation}
Here, $\otimes$ denotes a convolution and $\delta$ the delta function. Both sidebands contain the same information because they are complex conjugates of each other. Therefore, only a single sideband has to be extracted. The interference term can be extracted by masking the sideband of choice and applying a forward FT,
\begin{equation}
\begin{split}
I_+ &= \ft{ M_+ \ift{I}}  \\
&= A_s A_r^\dagger \exp{\{\frac{2\pi}{\lambda} i \alpha \epsilon \}}.
\end{split}
\end{equation}
This still contains the reference electric field; however, because the reference electric field is well known, it can easily be divided out. The reference electric field is the FT of a pinhole with a correction for the offset. The stellar electric field can then be reconstructed as
\begin{equation}
\begin{split}
A_s \approx \frac{I_+}{A_r^\dagger\exp{\{\frac{2\pi}{\lambda} i \alpha \epsilon \}}}.
\end{split}
\label{eq:ase}
\end{equation}
The reconstructed speckles, $A_s$, are the coronagraphic residuals, which still include the effects of the coronagraphic focal-plane mask. The incoming electric field can be reconstructed by dividing the estimate of $A_s$ by the focal-plane mask function \citep{mazoyer2013labverification}. However, this only works well for phase masks and does not work well for opaque focal-plane masks that remove light and therefore information. Fortunately, there is still a linear relation between the pupil plane electric field and the coronagraphic speckles, so it is still possible to build an interaction matrix with a deformable mirror (DM) to estimate and control the speckles.

\section{Modulating a self-coherent camera}
The SM-SCC has essentially the same layout as an SCC, with the only difference being that the pinhole has a spectral filter on top of it and that each spectral channel is imaged independently. This can be achieved by either using a chromatic beam splitter, which is usually used for DBI, or by using an IFS.

\subsection{Spectrally modulated self-coherent camera theory}
The SM-SCC can mimic the behavior of the FMSCC by blocking the pinhole of one of the spectral channels of a DBI setup. In that case, we get two focal-plane images,
\begin{equation}
I_1 = |A_{s,1}|^2 + I_{\mathrm{inc, 1}} 
\end{equation}
and
\begin{equation}
I_2 = |A_{s,2}^2|^2 + I_{\mathrm{inc,2}^2} + |A_{r,2}|^2 + 2 \Re\left\{A_{s,2} A_{r,2}^\dagger \exp{\left(\frac{2\pi i \vec{\alpha}\cdot\vec{\epsilon}}{\lambda_2}\right)}\right\}.
\end{equation}
Here, the additional index indicates the wavelength index, so $A_{s,1}$ is the stellar focal-plane electric field in wavelength channel 1. Following Eq. \ref{eq:otf}, the OTF of the first channel is
\begin{equation}
\begin{split}
\ift{I_1} &= \ift{|A_{s,1}|^2 + I_{\mathrm{inc},1}},
\end{split}
\label{eq:otf_wave1}
\end{equation}
and the OTF for the second image is
\begin{equation}
\begin{split}
\begin{split}
\ift{I_2} &= \ift{|A_{s,2}|^2 + I_{\mathrm{inc},2} + |A_{r,2}|^2} \\
&+ \ift{A_{s,2}^{\dagger} A_{r,2}}\otimes \delta(r - \epsilon) \\
&+ \ift{A_{s,2} A_{r,2}^\dagger}\otimes \delta(r + \epsilon).
\end{split}
\end{split}
\label{eq:otf_wave2}
\end{equation}

\begin{figure}
        \centering
        \includegraphics[width=\columnwidth]{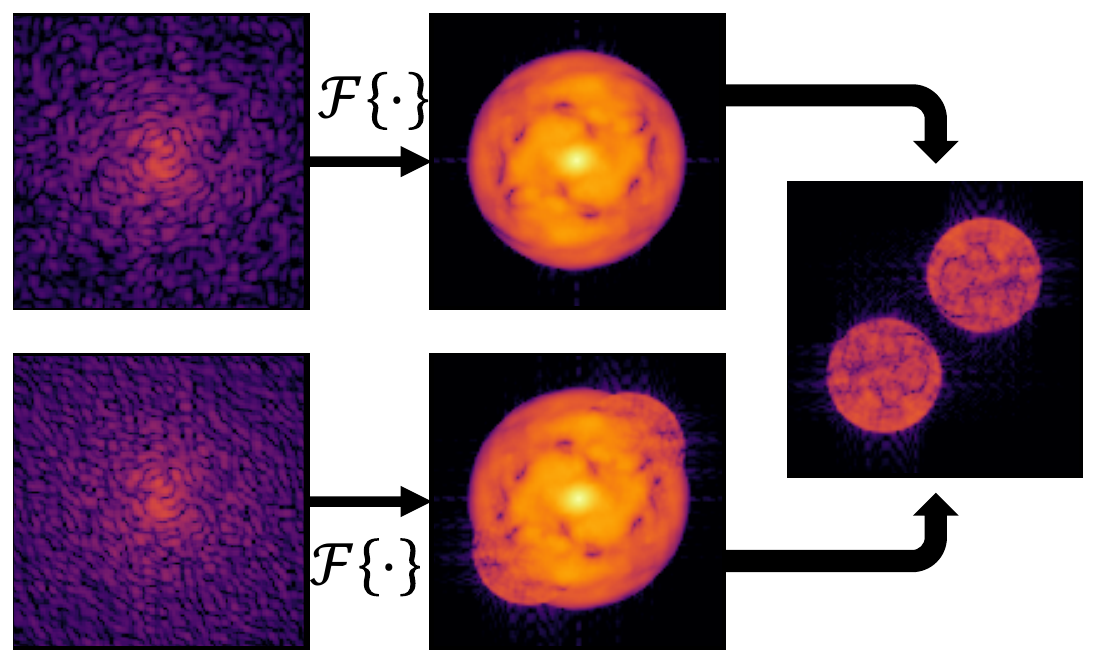}
        \caption{Post-processing flow of the sideband extraction process using two spectral channels. One channel has the pinhole blocked, while it is transparent for the other channel. The focal-plane images are Fourier-transformed to make the OTFs. Both contain the central OTF, while only one contains the sidebands. Subtracting the OTF of channel 1 from that of channel 2 reveals the sidebands. \edited{The OTF of the pinhole itself has also been subtracted to highlight the sidebands. The actual sensing process does not remove the pinhole OTF.}}
        \label{fig:otf_extraction}
\end{figure}
To extract the sidebands, the central OTF needs to be removed. The sideband can be masked if the sideband is separated enough from the central OTF, which is similar to the analysis in Sect. \ref{sec:wfs}. For the SM-SCC, another option is to process the two independent images: One image can be subtracted from the other to remove the central OTF. This is the approach that is used for the FMSCC. However, it is not possible to directly subtract the OTFs from the two different channels because the stellar speckles are chromatic. There are two scaling factors that need to be taken into account. First, due to the natural scaling of diffraction with wavelength, the speckles grow in size with wavelength. This has to be compensated for by multiplying the second OTF by $(\lambda_2 / \lambda_1)^2$. The second scaling factor comes from the chromatic speckle strength. In the small aberration regime with only phase errors, the incoming wavefront can be written as
\begin{equation}
E = P\exp{i\phi}\approx P\left(1 + i\phi\right).
\end{equation}
Here, $P$ is the pupil function and $\phi$ the phase errors. A perfect coronagraph will remove the constant pupil function \citep{cavarroc2006fundamental} and create the coronagraphic electric field,
\begin{equation}
E_c = iP\phi.
\end{equation}
Most atmospheric and NCPA speckles are created by optical path difference (OPD) errors. The relation between the phase and the OPD is $\phi=2\pi W / \lambda$. This leads to the coronagraphic residual,
\begin{equation}
E_c = i \frac{2\pi}{\lambda} P W.
\label{eq:ec_opd}
\end{equation}
The speckle intensity scales with $\lambda^{-2}$. Before subtracting the OTF of channel 1 from that of channel 2, the OTFs have to be compensate for both the chromatic scaling due to diffraction and the chromatic speckle strength, which results in a scaling of $(\lambda_2 / \lambda_1)^4$. Of course, this assumes that the spectrum of the star is equal in both channels. Another option is to use the OTFs themselves to compensate for a difference in flux between the channels due to the star. \edited{Not every part of the OTF can be used to estimate the relative flux of the object. The center of the OTF, for example its peak, has an additional contribution from the OTF of the pinhole. This will bias the measurement of the relative flux between the spectral channels. However, only a small part of the OTF has a contribution of the pinhole OTF because the pinhole is much smaller than the pupil diameter. Therefore, the parts of the central OTF that do not contain the pinhole OTF or the sidebands can be used to estimate the flux.} After compensation, channel 1 can be subtracted from channel 2,
\begin{equation}
\begin{split}
\begin{split}
\ift{I_2} - \alpha\ift{I_1} &\approx \ift{|A_{r,2}|^2} \\
&+ \ift{A_s^{2 \dagger} A_R}\otimes \delta(r - \epsilon) \\
&+ \ift{A_s^2 A_R^\dagger}\otimes \delta(r + \epsilon).
\end{split}
\end{split}
\label{eq:otf_multi_waves}
\end{equation}
Here, $\alpha$ is the chromatic correction factor, which is equal to $\alpha=(\lambda_2 / \lambda_1)^4 S_1 / S_2$, with $S_1$ and $S_2$ the intrinsic stellar flux in the two channels. The OTF of the reference hole has a diameter of $2d_p$. After removing the central OTF, one of the sidebands can be masked and the same procedure can be followed as for a normal SCC. Figure \ref{fig:otf_extraction} shows an example of the extraction procedure.

\subsection{SM-SCC pinhole separation and throughput}

\begin{figure}
        \centering
        \includegraphics[width=\columnwidth]{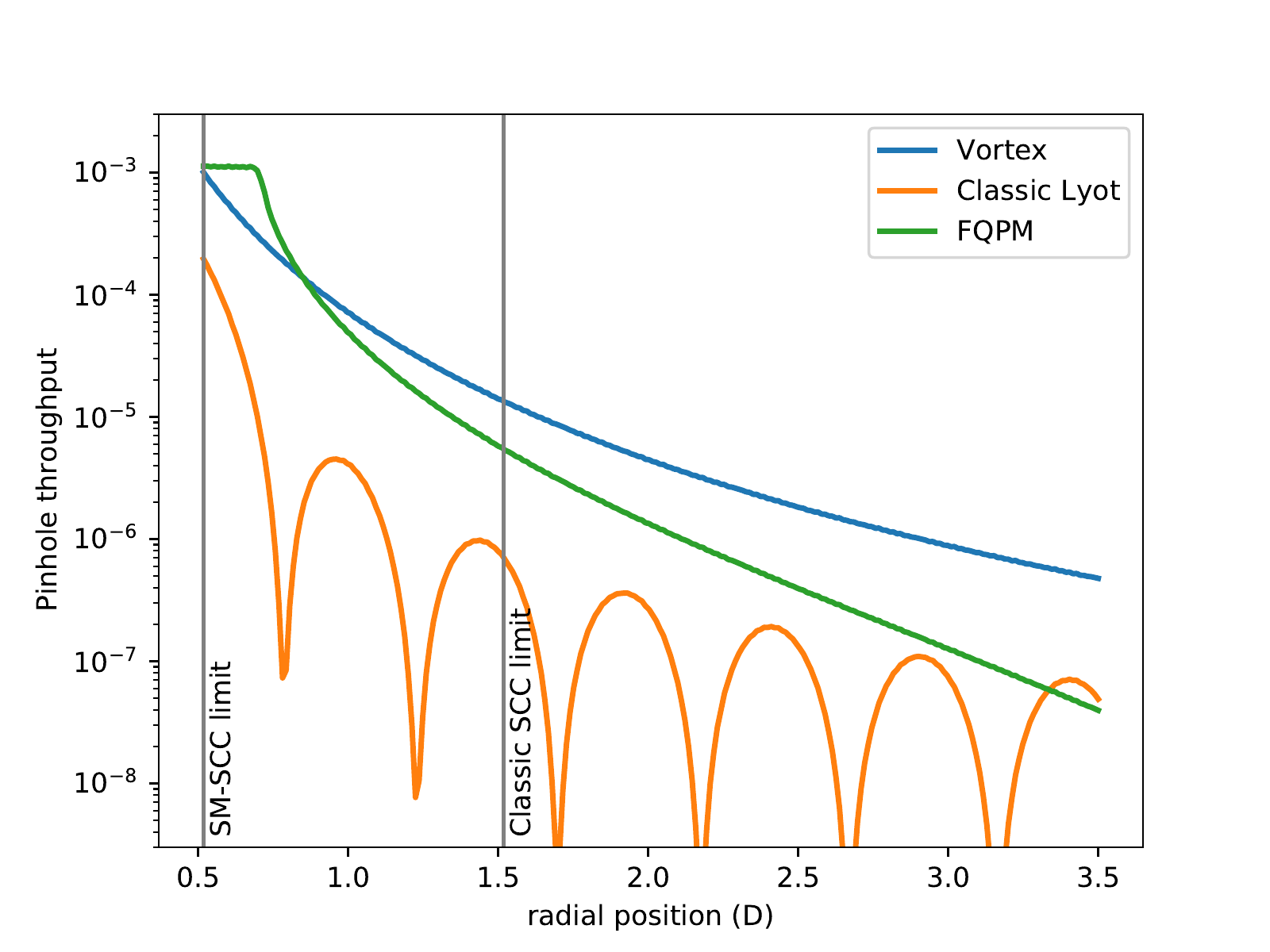}
        \caption{Throughput as a function of radial position for a $D/50$ pinhole for three different coronagraphs. Both the vortex coronagraph and the four quadrant phase mask coronagraph have higher throughput than a \edited{Lyot coronagraph with a 2 $\lambda/D$ circular opaque mask}, which shows that the phase mask coronagraphs with small inner working angles are preferred. Both the SM-SCC and the classic SCC limits are shown. The SM-SCC throughput is almost two orders of magnitude larger than the SCC limit.}
        \label{fig:coronagraph_throughput}
\end{figure}

The spectral modulation is used to separate the interference term from the central OTF, and there is no longer any need to create a high-frequency fringe. Therefore, the SM-SCC can place the pinhole closer to the edge of the pupil. \edited{The requirements for the distance are that the sidebands do not overlap with each other and that the sidebands do not overlap with the pinhole OTF}; these requirements result in the minimum distance $(\epsilon_m)$ of
\begin{equation}
\epsilon_m = \frac{1}{2}\left(1 + 2\gamma\right)D.
\label{eq:mod_pinhole_distance}
\end{equation}
With this separation, the throughput can be increased significantly. In the context of the SCC, the throughput is defined as the fraction of starlight that passes through the reference pinhole. Figure \ref{fig:coronagraph_throughput} shows the throughput for a $D/50$ pinhole as a function of radial position for three coronagraphs. In all three cases, the pinhole is moved across the diagonal direction. Both the    four-quadrant phase mask and the vortex coronagraph have a higher throughput than the classic Lyot coronagraph. For all three coronagraphs, the throughput of the SM-SCC can be 100 times higher than that of the SCC. 
\edited{The signal-to-noise ratio (S/N) of the SCC is complex because it depends on the ratio between the speckle intensity and the reference field \citep{galicher2010scc}. In the case where the speckle field is much stronger than the reference field, the S/N is limited by the power in the reference beam. In that case, the S/N is $\propto \sqrt{I_r}$. For a classic SCC, the reference field is on the order of $10^{-9}$ to a few times $10^{-8}$. Most instrumental speckles, and even residual AO speckles, are at a contrast of $10^{-6}$ to $10^{-4}$. The power in the reference is the limiting factor of the S/N for ground-based instruments. Increasing the throughput of the pinhole will increase the S/N of the wavefront measurements.}

In theory, the SM-SCC could gain a factor of $\sim \sqrt{100}=10$ in S/N. However, one of the channels does not contain the interference fringe. The channel without the fringe is used to remove the central science OTF and contributes additional noise. This effectively decreases the S/N by $\sqrt{2}$, leading to a total gain of $\sim 7$. However, this will still increase the limiting magnitude of the SCC by $2.5 \log_{10}{7^2} \approx 4$ magnitudes.

\subsection{Sampling requirements}

Adequate sampling of the fringes in the focal plane is required to resolve them. The minimum sampling is determined by the diameter of the OTF for band-limited functions. For a classic SCC, the maximal extent is $r_\mathrm{max} = \epsilon_0 + (1+\gamma)D/2$, which translates into a diameter of $2\epsilon_0 + (1 + \gamma)D = (4+2\gamma)D$. In the limit of a vanishing pinhole, $\gamma \rightarrow 0$, the diameter of the OTF becomes $4D$, which means that a minimal sampling of $\lambda / 4D$ per pixel is needed to retain the information from all frequencies.

For a modulated SCC, the extent of the OTF is $2\epsilon_m + (1+\gamma)D = (2+\gamma)D$. In the vanishing pinhole limit, we again arrive at a total diameter of $2D$. \editedn{Therefore, the focal plane only needs to be sampled at $\frac{\lambda}{2D}$ per pixel.} The modulated SCC reduces the sampling by a factor of two compared to a classic SCC. This reduces the influence of detector noise and allows the SM-SCC to work on fainter objects. More importantly, because the pixel sampling is reduced, an SCC can be used by IFSs for which it is difficult to create sufficiently high oversampling due to the limited detector real estate.

\section{Dual-band imaging simulations}
In this section the properties of the SM-SCC for a DBI system are explored through numerical simulations. All simulations are performed with the High Contrast Imaging for Python (HCIPy) package \citep{por2018hcipy}. The effects of photon noise and the spectral separation between the two channels on the quality of the wavefront reconstruction are the main focus. Earlier work has shown that the spectral bandwidth is important due to the chromatic smearing of the fringes \citep{galicher2010scc,mazoyer2014broadbandscc,delorme2016focal}. However, this is not relevant for the DBI observing mode. Most DBI modes use relatively narrow filters; for example, the Magellan Adaptive Optics eXtreme (MagAO-X) system has an H$\alpha$ DBI mode with a 1.25\,\% spectral bandpass \citep{males2018magaox}. \citet{galicher2010scc} derived the minimal spectral resolution that still allows the SCC to work as $R=N_{\mathrm{act}}\sqrt{2}\epsilon_0 / D$. For a 40-actuator DM and a separation of $\epsilon_0=1/2 D$, the minimal required resolving power is $R\approx28$. The spectral bandwidth that corresponds to this resolving power is $\Delta \lambda / \lambda = 3.6\,\%$. The widths of the MagAO-X filters are significantly smaller than this limit, which is why it is assumed here that the effects of spectral smearing can be neglected.

The simulated DBI system uses two spectral channels, with the central wavelength of the first channel fixed to 1 $\mathrm{\upmu}$m. The wavelength of the second channel is varied to investigate the effect of the small-phase approximation in Eq. \ref{eq:ec_opd}. The optical system has a clear aperture followed by a 40x40 DM, a vortex coronagraph, and \edited{a 5 percent undersized} LS with a pinhole covered by a spectral filter. The pinhole has a size of $\gamma = 0.02$ and is placed at $\epsilon_0 = 0.545D$. The pinhole can, in theory, be placed closer to the pupil edge, but placing it slightly farther out helps to suppress numerical artifacts in the wavefront reconstruction. \edited{Each channel in DBI observations is usually close to monochromatic, which is also assumed for the simulations in this section.}

\subsection{Wavefront sensing}

\subsubsection{Photon noise influence}
\begin{figure}
        \centering
        \includegraphics[width=\columnwidth]{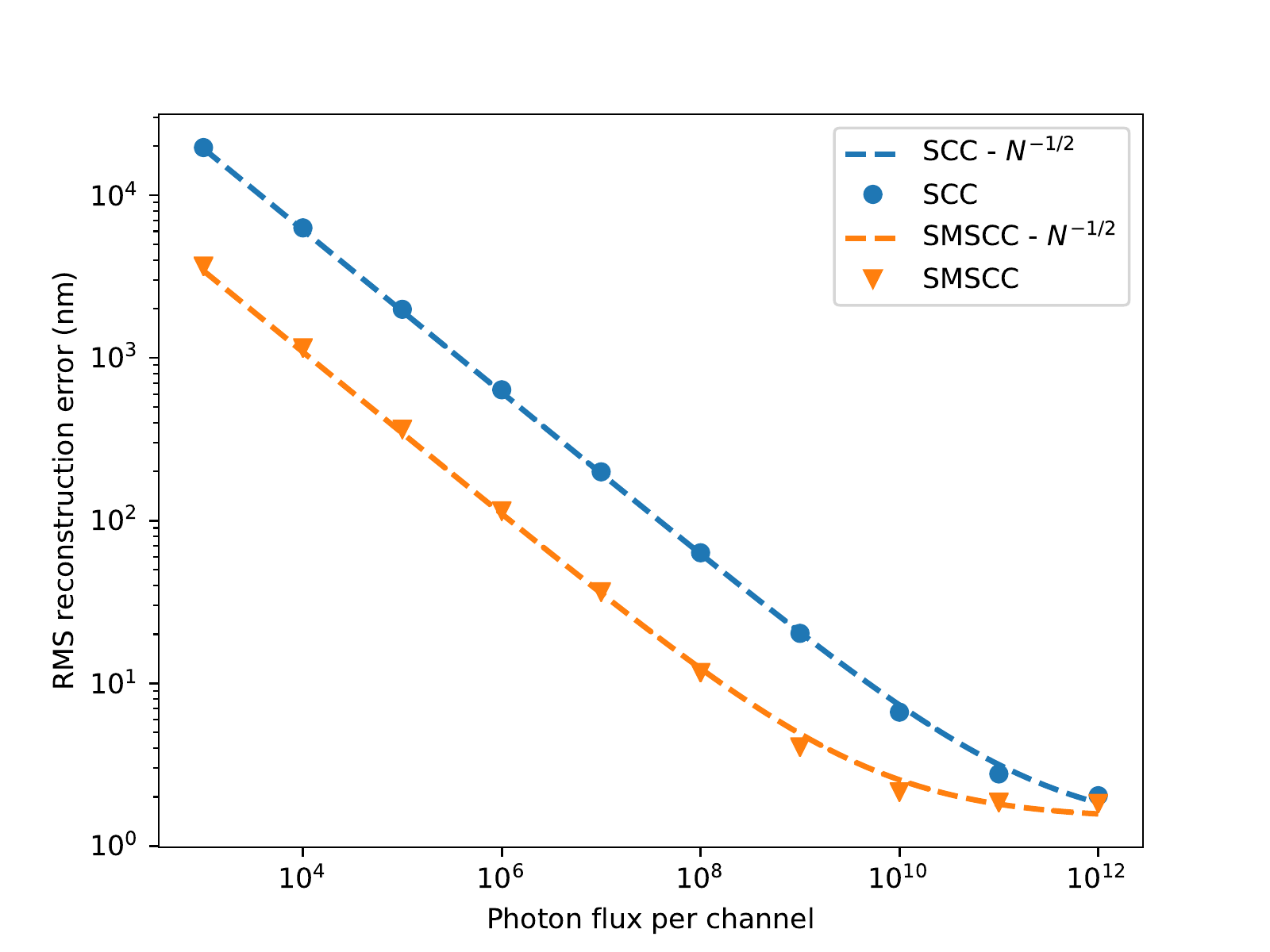}
        \caption{Reconstruction rms as a function of photon flux per spectral channel. The blue circles show the rms of a classic SCC, and the orange triangles show the rms of the SM-SCC. The dashed lines are fits of the expected $N^{-1/2}$ behavior due to photon noise. Around a photon flux of $10^9-10^{10}$, the curve of the SM-SCC starts to flatten out due to intrinsic reconstruction errors. For a classic SCC, this happens at a higher photon flux but at a similar reconstruction rms. The input rms was 30 nm, which indicates that both sensors can reconstruct the wavefront with errors smaller than 5\% at high flux levels.}
        \label{fig:photon_noise}
\end{figure}

The main benefit of the SM-SCC is that the reference hole can be placed closer to the edge of the pupil. This increases the flux and should increase the sensitivity. As shown in Fig. \ref{fig:coronagraph_throughput}, the number of photons is increased significantly by placing the pinhole closer to the geometric pupil edge. For the chosen pinhole parameters, the flux increases by a factor of 64, which should create a gain in sensitivity of $\sqrt{64/2}\approx5.66$. The sensitivity increase is estimated numerically by reconstructing the shape of the DM while the photon flux is varied. The DM was randomly deformed by adding 30 nm rms of white noise to the actuators. The reconstructed wavefront \edited{rms as a function} of photon flux for both a classic SCC and the SM-SCC is shown in Fig. \ref{fig:photon_noise}. The simulations show that the reconstruction rms levels out at high photon fluxes for both SCC techniques. A curve following $a/\sqrt{N} + b$, with $a$ the photon efficiency and $b$ the asymptotic rms, was fitted to both the SCC and the SM-SCC. The sensitivity gain is the ratio of the $a$ coefficients. The gain of the SM-SCC is $5.6\pm0.3$, which is consistent with the expected sensitivity gain based on the throughput. The asymptotic reconstruction rms is $1.23$ nm and $1.45$ nm for the classic SCC and the SM-SCC, respectively. Both techniques reconstruct the wavefront with a relative error smaller than 5 \% when no photon noise is present. \edited{These simulations were done without read noise. The SM-SCC used two images, as opposed to the SCC, which used one image. This makes the effective bandwidth of the SM-SCC twice as large. If both channels are used for the SCC, the gain in S/N of the SM-SCC will be lowered by $\sqrt{2}$ to $\sim4$. A way to regain this loss in S/N is to include multiple pinholes in the SM-SCC. The pinholes in the different channels can be placed in such a way that the sidebands of the different filters do not overlap. This is similar to the principle of the multi-reference SCC \citep{delorme2016focal}}.

\subsubsection{Spectral separation between channels}
\begin{figure}
        \centering
        \includegraphics[width=\columnwidth]{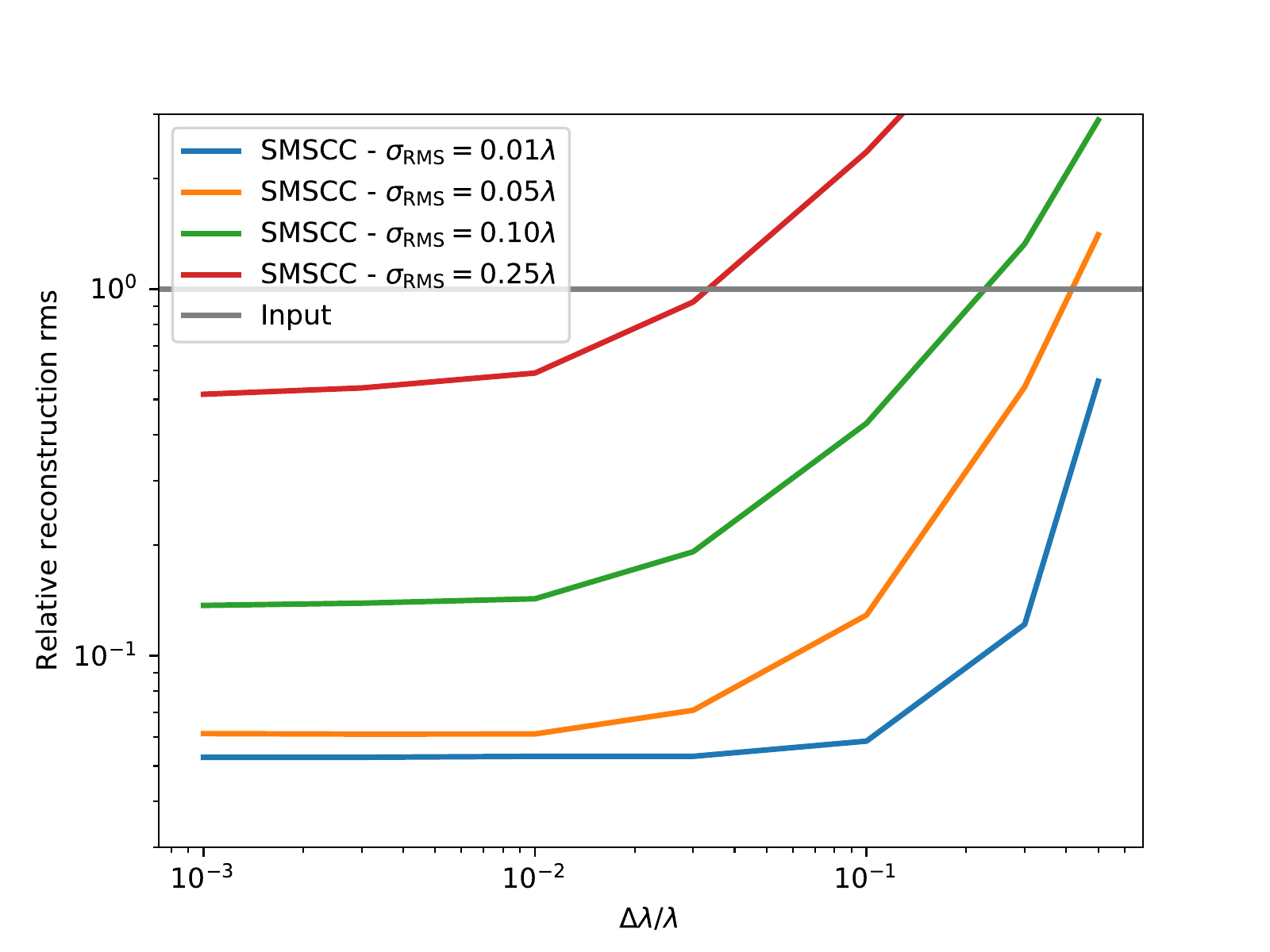}
        \caption{Relative reconstruction quality as a function of spectral separation for several levels of NCPA errors. The reconstruction quality is best for small spectral separations and small wavefront errors. The reconstruction quality converges at smaller spectral separations to the classic SCC quality. We see that the quality degrades when the NCPAs increase in strength, even for small spectral separations. The quality also degrades faster for large spectral separations than for small spectral separations. Both effects are due to higher-order terms that have been neglected in the small-phase approximation. A spectral separation smaller than 3\% always show improvement.}
        \label{fig:spectral_separation}
\end{figure}

The wavefront reconstruction quality depends on both the spectral separation between the two channels and the strength of the NCPAs. \edited{The spectral separation is defined as $\Delta \lambda = \lambda_2 - \lambda_1$}. In the previous simulations, the spectral separation was 1 \% between the two channels. This is representative for the H$\alpha$ mode of MagAO-X, but not for the DBI mode of SPHERE\citep{beuzit2019sphere}. The H2/H3 filters of SPHERE have a spectral separation of 4\%. To test the spectral channel limit, 100 \edited{random DM shapes were generated by adding Gaussian noise to the actuators} for each spectral separation and NCPA strength level. The wavefront reconstruction quality, $Q$, is measured as the ratio between the \edited{rms of the residual phase} and the input rms, \edited{Q = $\sqrt{||\phi_{\mathrm{in}} - \phi_{\mathrm{reconstructed}}||^2 / ||\phi_{\mathrm{in}}||^2}$}. \edited{Here, $||\cdot||$ is used as the norm of the vector.} When the quality is higher than 1, the estimated wavefront error is actually worse than the input wavefront error. The results are shown in Fig. \ref{fig:spectral_separation}. A shorter separation in wavelength between the two channels reduces the \edited{residual} rms. The reconstruction quality is best for small spectral separations and small wavefront errors. The reconstruction quality converges for smaller spectral separations to the classic SCC quality. The quality degrades when the NCPA increases in strength, even for small spectral separations. The quality also degrades faster for large spectral separations than for small spectral separations. Both effects are due to higher-order terms that have been neglected in the small-phase approximation. The higher-order terms have a different spectral behavior. If different orders play a role for different speckles, the wavelength scaling compensation will be different. This makes it difficult to apply the correct wavelength scaling to remove the central OTF. In general, the wavefront rms is smaller than 50 nm on bright stars. For the H2/H3 DBI mode of SPHERE, this corresponds to a wavefront rms of $0.03 \lambda$ \citep{n2013calibration} at a separation of 4\%. The H$\alpha$ DBI mode on MagAO-X has a smaller spectral separation of 1\%, with an expected residual NCPA level of <30 nm rms or $<0.05\lambda$ \citep[][Lumbres et al. in prep]{van2021characterizing}. Both \edited{systems} can easily be handled by the SM-SCC, according to Fig. \ref{fig:spectral_separation}.

\subsubsection{Differential aberrations between channels}
\begin{figure}
        \centering
        \includegraphics[width=\columnwidth]{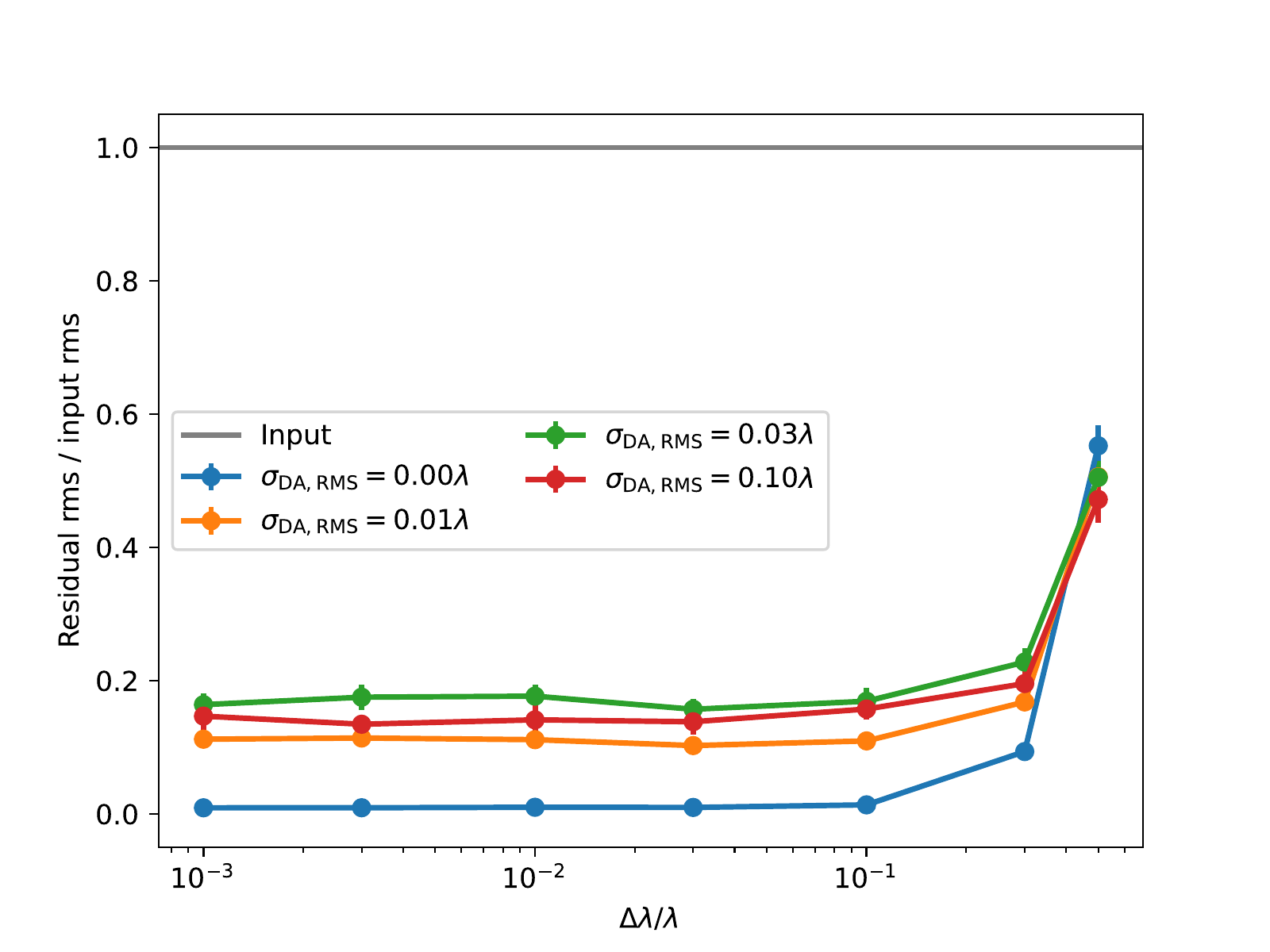}
        \caption{\edited{Normalized residual rms} as a function of spectral separation for several levels of DAs. The reconstruction quality is best for small spectral separations and small DAs. The quality degrades when the DAs increase in strength, even for small spectral separations. The degradation is independent for a spectral separation smaller than 10 percent.}
        \label{fig:diff_aberration}
\end{figure}

In DBI there is usually a beam splitter somewhere in the optical path to separate the two spectral channels. The beam splitter and all subsequent optics will create differential aberrations (DAs) between the two spectral channels. This will have an impact on the reconstruction quality of the wavefront. Most DBI setups try to minimize such DAs. For example, the DBI mode of SPHERE/IRDIS requires that the DAs be smaller than 10 nm rms, with a goal of 5 nm rms \citep{dohlen2008infra}. The impact of the DAs was tested by adding NCPA errors to only one of the two channels. \edited{There are two different cases that have to be considered. For the first one, the DAs are static and therefore can be included in the calibration. In the second case, the DAs are dynamic, which means they cannot be calibrated. For the first case, the aberrations are included in the system and the SM-SCC} is then calibrated to partially compensate for the influence that the aberrations have. The reconstruction quality was again determined by generating 100 random DM states and measuring the rms of the \edited{residual phase} relative to the input rms. The simulation results in Fig. \ref{fig:diff_aberration} show that increasing the amount of DAs does decrease the reconstruction quality. However, there is no strong dependence on the spectral separation because the sensor was calibrated with the NCPAs included.

With the DA strength of SPHERE/IRDIS, the performance will lie between the curves of \edited{$0.01\lambda$ and $0 \lambda$ rms}. These curves show a reconstruction quality of about 90\%, which is more than enough for closed-loop operations. These simulations demonstrate that static DAs do not influence the reconstruction quality of the SM-SCC if care is taken to minimize the DAs and if the aberrations are included in the sensor's calibration.

\begin{figure}
        \centering
        \includegraphics[width=\columnwidth]{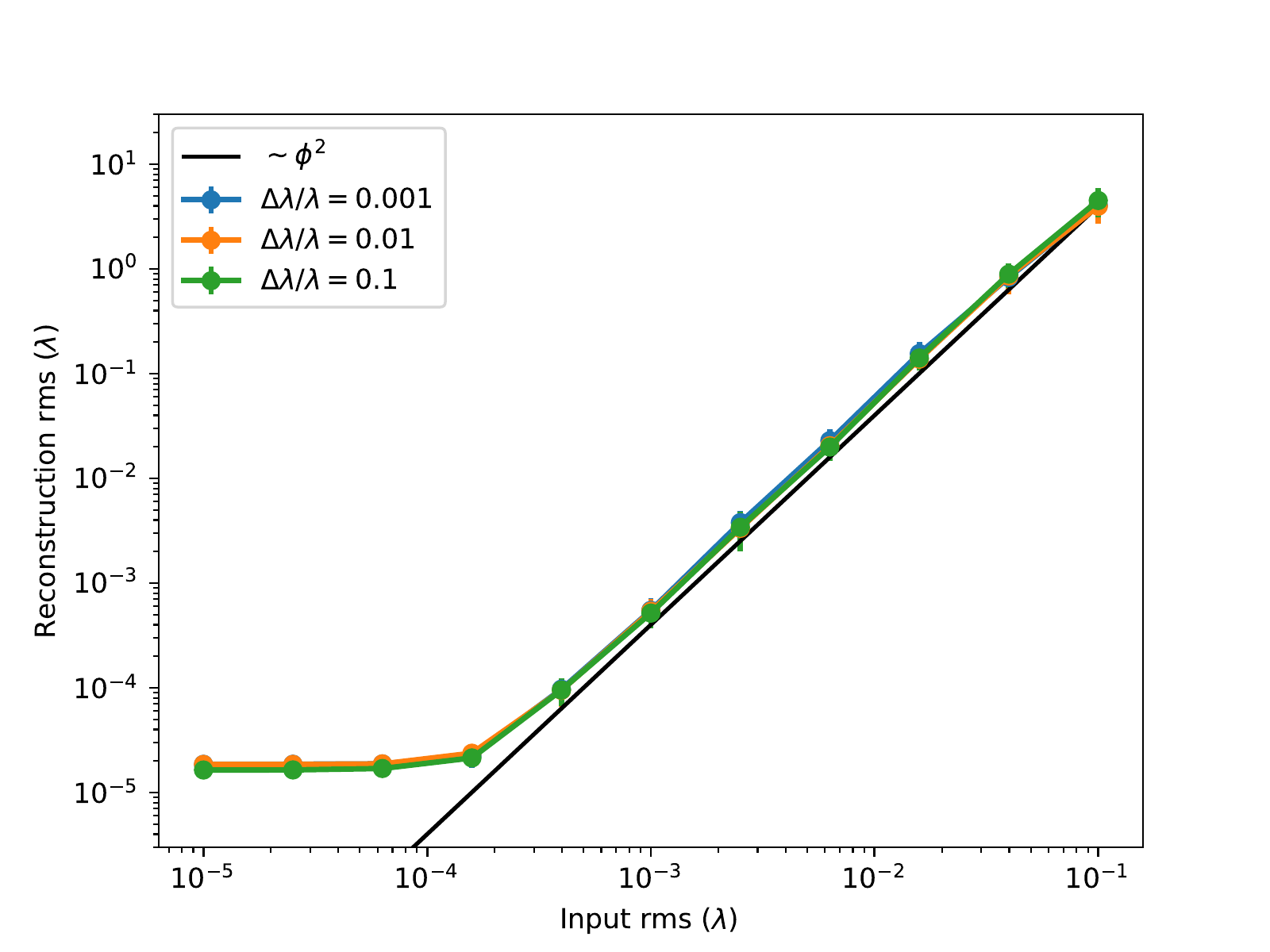}
        \caption{\edited{Reconstruction rms in waves as a function of uncalibrated DA rms. The different colors represent spectral separations. The reconstruction error closely follows a quadratic function, shown in black, which indicates that the DAs come in as a second-order effect. The rms at small aberrations is limited by the reconstruction algorithm at $\sim10^{-5} \lambda$.}}
        \label{fig:diff_dynamic_aberration}
\end{figure}

\edited{More problematic are the dynamic DAs. The second channel, the one without the pinhole, is used to subtract the unmodulated speckles. Any non-common speckle will fold back into a reconstruction error. The effect on the reconstruction rms is shown in Fig. \ref{fig:diff_dynamic_aberration}. A reconstruction rms lower than $\lambda/100$ requires a DA rms smaller than $\lambda/200$. While this may look very strict, it is within current manufacturing capabilities. The DBI mode of SPHERE/IRDIS has a requirement of a less than 10 nm DA rms with a goal of 5 nm rms. This corresponds to $\lambda/320$ to $\lambda/160$ in the H1/H2 filters (1600\,nm). Additionally, this NCPA describes the total amount of DA and includes the dynamic and static DAs between the two channels.}

\edited{In this section the DAs are injected before the coronagraph. Typical beam splitters and DBI instruments split the light after the LS. This means that the DAs between the two paths are minimized. The DAs after the LS will slightly modify the speckles, but these errors are at a fraction of the speckle intensity themselves and are also proportional to the speckle strength. This means that when the speckles are removed in a closed loop, the errors will also go away. Therefore, the upstream aberrations will not play a major role for the SM-SCC. }

\subsection{Wavefront control}
In this subsection the performance of the SM-SCC with closed-loop feedback is investigated. The optical system under consideration has a clear aperture followed by a 40x40 DM, a vortex coronagraph, and an LS with a pinhole covered by a spectral filter. The pinhole has a size of $\gamma = 0.043$ and is placed at $\epsilon_0 = 0.568D$. Two different situations are simulated: a system with only phase aberrations and a system with phase and amplitude aberrations. In the first system all speckles within the control radius will be removed, while for the second a one-sided dark hole will be created. Not all speckles can be removed in the second case because both amplitude and phase aberrations are present and the system only has a single DM.

For both cases, the effects of photon noise and spectral separation between the two channels are considered. The amount of photon noise is based on the end-to-end flux expected by MagAO-X\footnote{https://magao-x.org/docs/handbook/observers/filters.html}. For MagAO-X, a zeroth magnitude star in a \editedn{narrow-band filter} has roughly $3 \cdot 10^9$ photons per second. Three different magnitudes were tested, 0th, 5th, and 7.5th. The corresponding photon fluxes are $3 \cdot 10^9$, $3 \cdot 10^7$, and $3 \cdot 10^6$ photons per second. \editedn{Only the correction of the slowly evolving quasi-static speckles is investigated in this work. Therefore, the integration time of the sensor is set to 1 second per frame.} The wavelength of channel 1 was fixed to $1 \upmu$m, while the wavelength of channel 2 was varied. The considered spectral separations were $\Delta \lambda/\lambda = 0.01$, $\Delta \lambda/\lambda = 0.05,$ and $\Delta \lambda/\lambda = 0.1$.

\subsubsection{Full dark hole wavefront control}
\editedn{In the phase-only case, the full control region is controlled. The phase aberration that is injected as NCPA has a power spectral density (PSD) that follows a $f^{-2.5}$ power law with a total of 30 nm rms, which is typical for instrumental aberrations. The control region is a square area of $40 \lambda / D$ by $40 \lambda / D$. The control matrix is built up empirically by applying random patterns to the DM and measuring the response in the extracted side band. If enough random patterns are collected the reconstruction matrix, or control matrix, can be calculated as}
\begin{equation}
C = V I_{\mathrm{im}}^T \left(I_{\mathrm{im}} I_{\mathrm{im}}^T + \mu I \right)^{-1}
\end{equation}
The reconstruction matrix, $C$, depends on both the input probes, $V$, and the recorded images, $I_{\mathrm{im}}$. The identity matrix, $I$, is used as a regularization matrix, with $\mu$ the strength of the regularization. \edited{The regularization parameter, $\mu$, was chosen by minimizing the reconstruction error on a subset of the measurements. This can be done if the number of measurements is larger than the number of DM modes. The subset contained 10 percent of the total data.} A leaky integral controller was used to control the DM. For the high flux cases (0th and 5th magnitude) the gain was 0.5, while for the low flux case (7.5th magnitude) a gain of 0.25 was used.

\begin{figure*}
        \centering
        \includegraphics[width=\textwidth]{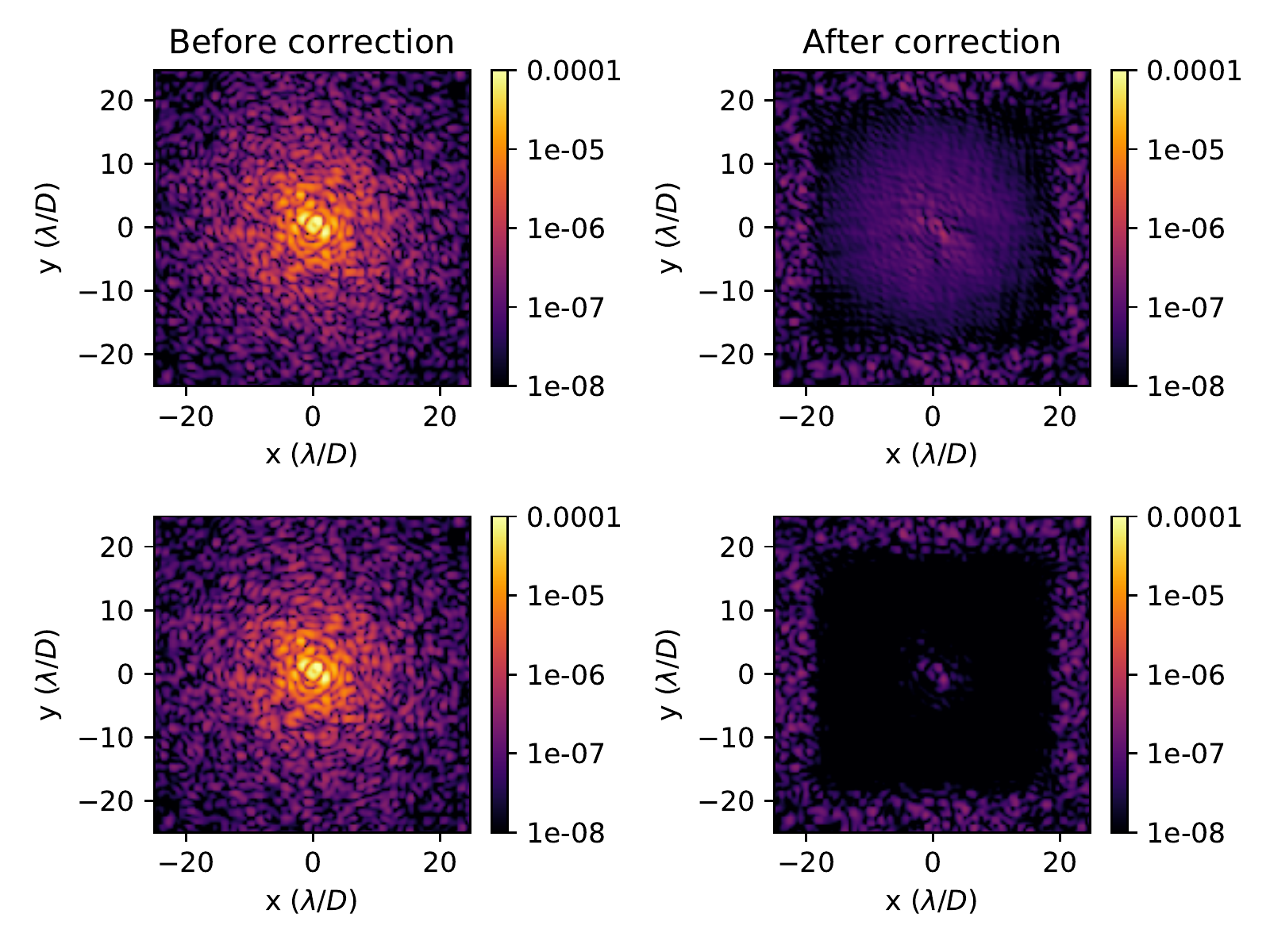}
        \caption{Focal-plane contrast before and after wavefront control for a zeroth magnitude star. The images in the top row are from the channel that has light from the pinhole, while the images on the bottom are from the channel with the blocked pinhole. The wavefront aberrations of this simulation are phase only (75 nm ptv), which allows for control of the full control region with a single DM. The control loop ran for 25 iterations, and the final dark hole contrast is well below $10^{-8}$ for channel 2. The contrast in the dark hole of channel 1 is limited by the light from the pinhole. }
        \label{fig:bright_star}
\end{figure*}

\begin{figure*}
        \centering
        \includegraphics[width=\textwidth]{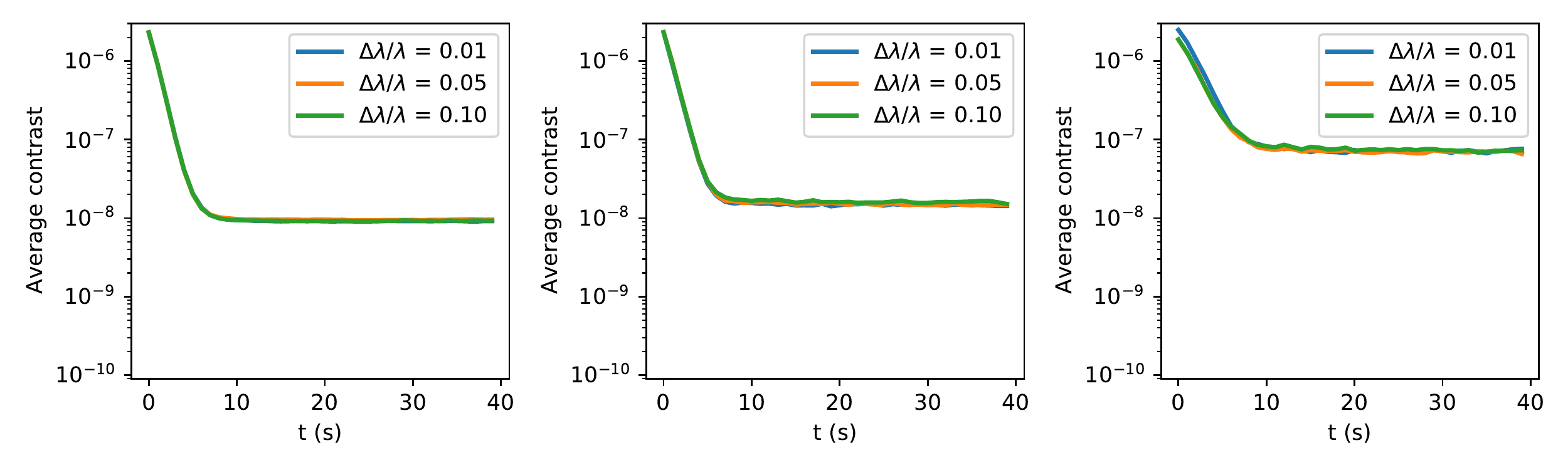}
        \caption{Median contrast in the dark hole region (-20 $\lambda/D$ to 20 $\lambda/D$ in x by -20 $\lambda/D$ to 20 $\lambda/D$ in y) as a function of time. Each frame has an exposure time of 1 second. The contrast is shown for three different spectral separations, which are shown in different colors. The three figures are for a 0th (left), 5th (middle), and 7.5th (right) magnitude star as observed by MagAO-X. The final median contrast is almost independent of the spectral separation between the two channels. The smallest spectral separation does lead to the deepest contrasts. The simulations with a fainter star (right) have been done with a lower feedback gain, which increased the convergence time to 25 s.}
        \label{fig:fullregion_contrast}
\end{figure*}

A before-and-after wavefront control snapshot is shown in Fig. \ref{fig:bright_star}. Channel 1 contains the reference electric field, which becomes visible after the wavefront control has been used to clean up the wavefront errors. Channel 2, without the reference field, reaches a median contrast of $1\cdot 10^{-8}$ over the full control region. At small separations (1 to 2 $\lambda/D$) there are a couple speckles left over that are at a contrast level of $1\cdot10^{-7}$. Figure \ref{fig:fullregion_contrast} shows the median contrast in the dark hole region as a function of time for different levels of photon flux and spectral separation between the two channels. The SM-SCC converges within ten iteration in the high flux case, with a gain of 0.5. For the low flux case the SM-SCC converges after 25 iterations. The expected degradation between the 5th and 7.5th magnitude simulations, based on photon noise statistics, is $\sqrt{10}\approx3$, which is also approximately what is seen. There seems to be little influence of the spectral separation on the achieved contrast in closed-loop feedback. During closed-loop feedback, the wavefront errors decrease, which reduces the effects of the higher-order terms. This in turn improves the small-phase approximation that was made for the control.

\subsubsection{One-sided dark hole wavefront control}
The controller for the case with both phase and amplitude errors is more complicated because a one-sided dark hole has to be created. \edited{The NCPA phase PSD follows an $f^{-2.5}$ power law with a total of 30 nm rms. The amplitude aberrations follow an $f^{-1.5}$ power law with a 20\% peak-to-valley.}

 In most cases, before the one-sided dark hole was created, the Strehl was increased by applying multiple steps of feedback with the full dark hole \citep{potier2020efcsphere}. This step was used to increase the Strehl of the system. After the full dark hole was corrected, the one-sided dark hole was created. This approach works well if the aberrations are static. However, there could be aberrations that need the full dark hole to be measured. If these modes are time variable, the Strehl will degrade even though the one-sided dark hole stays dark. This effectively decreases the contrast and throughput. This can be circumvented by using focal-plane weight maps where the pixels in the dark hole have a stronger weight. The weight maps were used to determine how much priority has to be given to create the dark hole and retain the Strehl. Setting the weight outside the dark hole to zero but inside the dark hole to 1 will ignore the response outside the dark hole. The controller is created in two steps. First, the image response matrix is generated,
\begin{equation}
R = I_{\mathrm{im}} V^T \left(V V^T + \alpha I \right)^{-1}.
\end{equation}
The response matrix, $R$, is the matrix that tells us how the focal plane responds to DM patterns. \edited{The regularization parameter, $\alpha$, was chosen in a similar way as the full dark hole control, by optimizing it on 10 percent of the calibration data.} A weighted least squares (WLS) system can then be set up to find the controller,
\begin{equation}
V_{\mathrm{opt}} = \argmin_V ||\Sigma^{1/2}\left(I_{\mathrm{m}} - RV\right)||^2 + \beta||V||^2.
\end{equation}
Here, $V_{\mathrm{opt}}$ is the optimal control command, $\Sigma$ the weight matrix, $I_{\mathrm{m}}$ the measurement, $V$ a control command, and $\beta$ a second regularization parameter. The solution to this problem is
\begin{equation}
V_{\mathrm{opt}} = \left(R^T \Sigma R + \beta I \right)^{-1} R^T \Sigma I_{\mathrm{m}}.
\end{equation}
The control matrix is then
\begin{equation}
C = \left(R^T \Sigma R + \beta I \right)^{-1} R^T \Sigma.
\end{equation}
For the simulations presented here, a weight of 1 was used within the dark hole, a weight of 0.01 within the control radius, and outside the control radius the weight was set to 0. The one-sided dark hole has a size of $15 \lambda/D \times 30 \lambda / D$, with its edge at $2.5 \lambda / D$. The control region is still $40 \lambda/D \times 40 \lambda / D$. \edited{The regularization parameter, $\beta$, was empirically chosen by finding the smallest value that resulted in converging closed-loop operations.}

\begin{figure*}
        \centering
        \includegraphics[width=\textwidth]{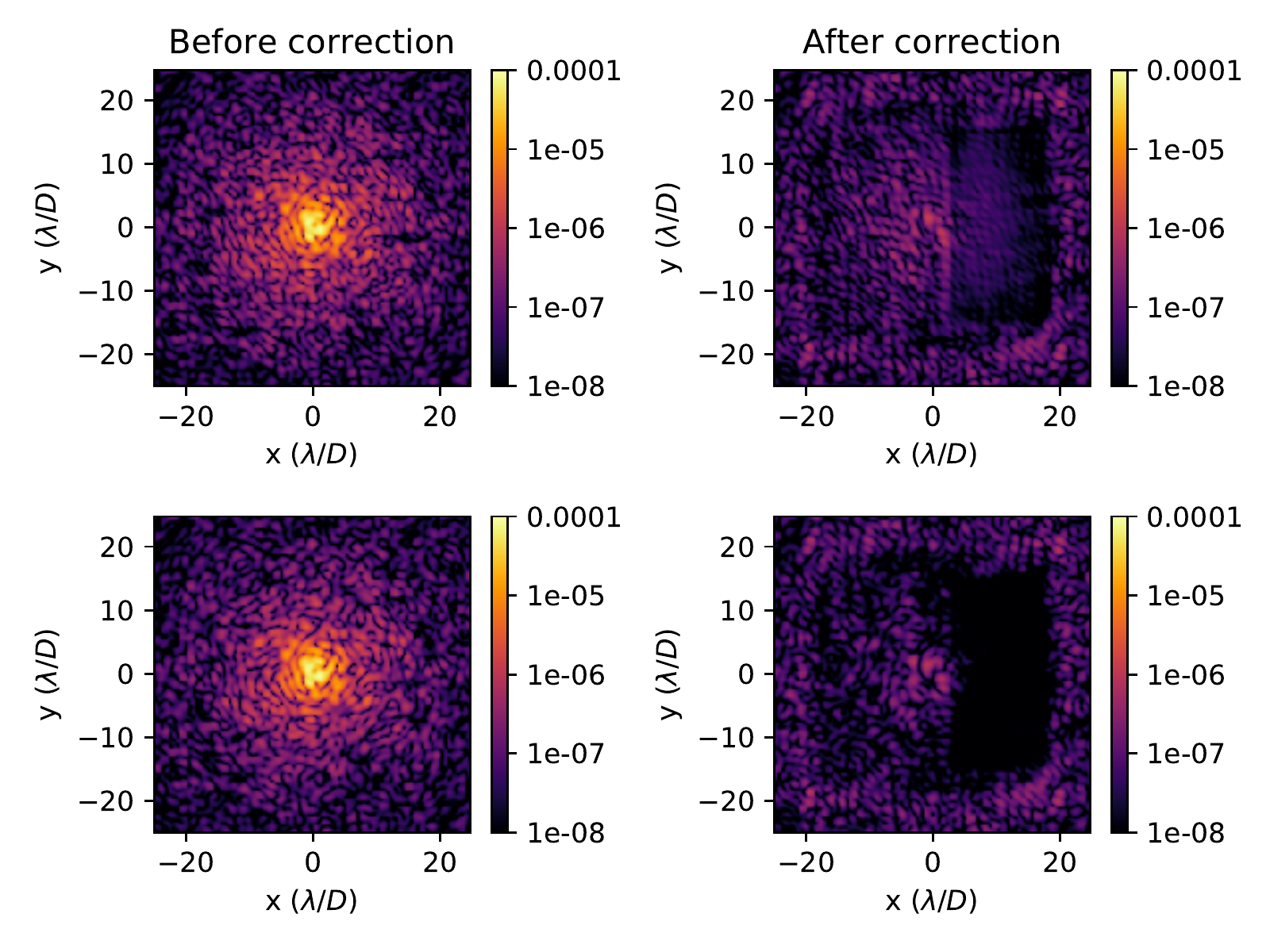}
        \caption{Focal-plane contrast before and after wavefront control for a zeroth magnitude star. The images in the top row are from the channel that has light from the pinhole, while the images on the bottom are from the channel with the blocked pinhole. This simulation contains both amplitude (20\,\%) and phase (75 nm ptv) aberrations. Only a one-sided dark hole can be created because there is only a single DM in the system. The control loop ran for 25 iterations, and the final dark hole contrast is well below $10^{-8}$ for channel 2. The contrast in the dark hole of channel 1 is limited by the light from the pinhole.}
        \label{fig:bright_star_with_amp}
\end{figure*}

\begin{figure*}
        \centering
        \includegraphics[width=\textwidth]{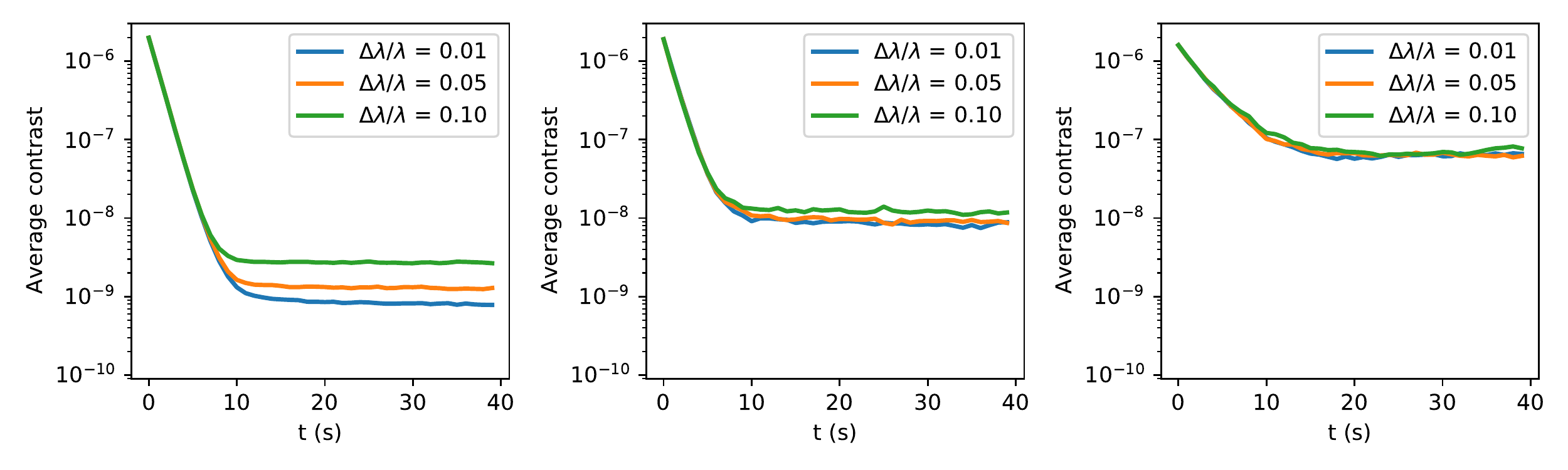}
        \caption{Median contrast in the dark hole region (2.5 $\lambda/D$ to 17.5 $\lambda/D$ in x by -15 $\lambda/D$ to 15 $\lambda/D$ in y) as a function of time. Each frame has an exposure time of 1 second. The contrast is shown for three different spectral separations, which are shown in different colors. The three figures are for a 0th (left), 5th (middle), and 7.5th (right) magnitude star as observed by MagAO-X. The final median contrast is independent of the spectral separation between the two channels.}
        \label{fig:onesided_contrast}
\end{figure*}

A before-and-after snapshot can be seen in Fig. \ref{fig:bright_star_with_amp}. This figure shows the impact of the amplitude aberrations. The contrast in the one-sided dark hole is deeper than $1\cdot 10^{-8}$, while outside the dark hole the contrast is between $1\cdot 10^{-7}$ and $1\cdot 10^{-6}$. Figure \ref{fig:onesided_contrast} shows the median contrast within the dark hole as a function of time. The one-sided dark hole reaches a contrast of $<1\cdot 10^{-9}$ for the smallest spectral separation. The contrast degrades as the wavelength difference is increased. Increasing the magnitude from the zeroth to the fifth increases the median contrast by a factor of ten, which is the expected degradation from photon noise. The difference between the three spectral separations is almost negligible, and photon noise dominates the residuals. At 7.5th magnitude, the contrast increases to $6\cdot10^{-8}$, which is more than the additional photon noise can explain. Both the regularization and feedback gain had to be adjusted to get the system to converge. A better result closer to the photon noise limit can probably be reached if both the regularization and control are optimized. These simulations show that the SM-SCC can correct both phase aberrations and amplitude aberrations. 
While there is no difference in the performance for the full dark hole when the spectral separation is changed, there is a \edited{small difference for the one-sided simulations at the brightest magnitude}. The main difference between these sets of simulations is the addition of amplitude errors. The speckles of amplitude errors grow differently with wavelength as compared to OPD speckles. The main chromatic effect of amplitude errors is the scaling due to diffraction $(\propto \lambda^{-2})$. It is not possible to differentiate between the two types of speckles in the DBI setup. This also hints at a limitation of the SM-SCC, the chromatic behavior of speckles. However, even with both phase and amplitude errors, a contrast limit of $<1\cdot10^{-7}$ can be reached for the faintest object. This is well below the current AO performance \citep{beuzit2019sphere, cantalloube2020windhalo}, and even below the predicted contrast limit on the next generation of telescopes, such as the Giant Magellan Telescope, that use advanced predictive control laws \citep{males2018lpc}.

The effects of residual atmospheric turbulence were neglected. The residual atmospheric turbulence will modulate the speckles and create an incoherent background. The ratio of the integration time and the coherence time will determine how strong the incoherent background will be. With the focal-plane feedback, the contrast in the dark hole will be limited to the incoherent background \citep{galicher2019sccpalomar, singh2019sccaohalo, potier2020efcsphere}. The exact effect that the residual turbulence will have on the SM-SCC will be part of a later work.


\section{Extending the SM-SCC to many spectral channels}
Instead of only two channels, as in the DBI mode, the SM-SCC can also use more channels when it is combined with an integral field unit (IFU). Most, if not all, direct imaging instruments contain an IFU for spectral characterization \citep{groff2015charis, beuzit2019sphere, sun2020hdfsluvoir}. Therefore, it is interesting to extend the SM-SCC to multiple channels. An added benefit of the multiple channels is the ability to use more complex modulation schemes. Not only can the pinhole be modulated, but the throughput can also be modulated through the LS. The residual stellar speckles, the reference probe, and the interference term can be reconstructed independently if a different modulation is applied to both the pinhole and the LS.

\subsection{Reconstruction theory}
If the spectral transmission of the LS is $a(\lambda)$ and the pinhole transmission is $b(\lambda)$, the intensity in the focal plane is
\begin{equation}
I = a^2|A_s|^2 + b^2|A_r|^2 + 2ab \Re\left\{A_s A_r^\dagger \right\}.
\end{equation}
Each of the three terms is modulated along the spectral direction at a different rate ($a^2$, $2ab$, and $b^2$). If the transmission coefficients $a$ and $b$ contain enough diversity, the three terms can be disentangled. For the reconstruction, all spectral channels have to be interpolated to the same wavelength by scaling the focal-plane image by a factor $\lambda / \lambda_{ref}$. For the second step, the intensities were scaled by $(\lambda / \lambda_{ref})^2$. The interpolation and scaling takes care of the chromatic scaling, as described in Sect. 3.1. A system of equations can be set up that relate the achromatized electric fields to the measured intensities,
\begin{equation}
\left[I_0 \cdots I_N\right] = \begin{bmatrix}
a(\lambda_0)^2 / \lambda_0^2 & 2ab / \lambda_0  & b(\lambda_0)^2 \\ 
\vdots & \vdots  & \vdots \\ 
a(\lambda_N)^2 / \lambda_N^2 & 2ab / \lambda_N  & b(\lambda_N)^2
\end{bmatrix}
\begin{bmatrix}
|A_s|^2\\ 
\Re \left\{ A_r^\dagger A_s \right\}\\ 
|A_r|^2
\end{bmatrix}.
\label{eq:system_equations}
\end{equation}
Here I have assumed that $A$ and $A_R$ are achromatic up to the chromatic scaling factor. This assumption is true for the \edited{Vector Vortex Coronagraph, which uses geometric phase optics \citep{murakami2013design}.} Inverting Eq. \ref{eq:system_equations} leads to
\begin{equation}
\begin{bmatrix}
|A_s|^2\\ 
\Re \left\{ A_R^\dagger A_s \right\}\\ 
|A_R|^2
\end{bmatrix}
= M \left[I_0 \cdots I_N\right].
\label{eq:system_equations2}
\end{equation}
The demodulation matrix, $M$, is the inverse of the spectral modulation matrix. The demodulation matrix allows the interference term, $\Re \left\{ A_r^\dagger A_s \right\}$, to be estimated. There is a linear relation between the interference term and the incoming wavefront. Therefore, the wavefront can be reconstructed, either from a linear model or an empirical interaction matrix. More importantly, the interference term can be reconstructed without explicit knowledge of the reference field. This adds the additional degree of freedom of LS engineering. An example would be adding more pinholes to increase the throughput.

\subsection{Demonstration of the multiband SM-SCC}
\begin{figure*}
        \centering
        \includegraphics[width=\textwidth]{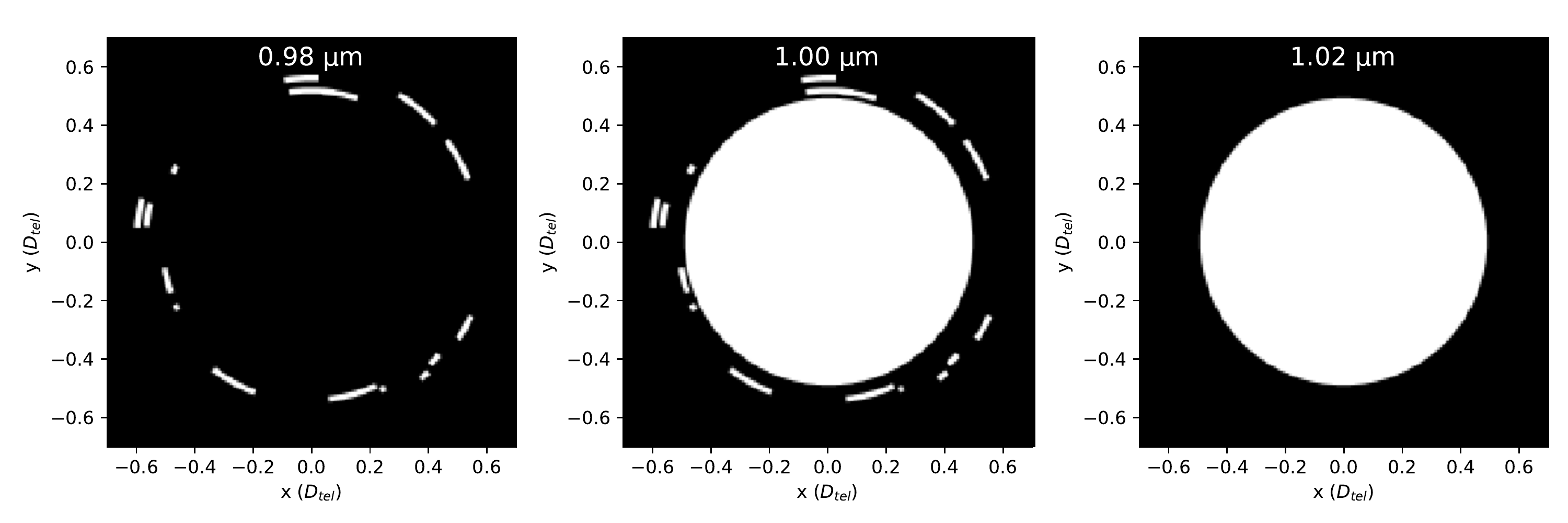}
        \caption{Modulated LS for the three different wavelength channels. Both the aperture and the reference field aperture are modulated.}
        \label{fig:multi_band_apertures}
\end{figure*}

A system with three spectral channels was used to showcase the added benefit of modulating both the LS and the pinhole. The reference channel for this simulation was not created with a pinhole, but with 16 random arcs. Each arc had an arc length that was drawn from a uniform distribution between $\pi/100 $ and $\pi/10$. The radius of the arc was also randomly drawn between $0.5 D_{tel}$ and $0.6 D_{tel}$. The width of each arc was $0.043 D_{tel}$, a similar width to the diameter of the pinholes from the previous section. The three spectral channels had a central wavelength of $0.98 \upmu$m, $1.0 \upmu$m, and $1.02 \upmu$m. A long-pass filter ($\geq1.0 \upmu$m) was chosen for the LS and a short-pass filter ($\leq1.0 \upmu$m) for the reference aperture. The LS for each of the three channels can be seen in Fig. \ref{fig:multi_band_apertures}. 

\begin{figure*}
        \centering
        \includegraphics[width=\textwidth]{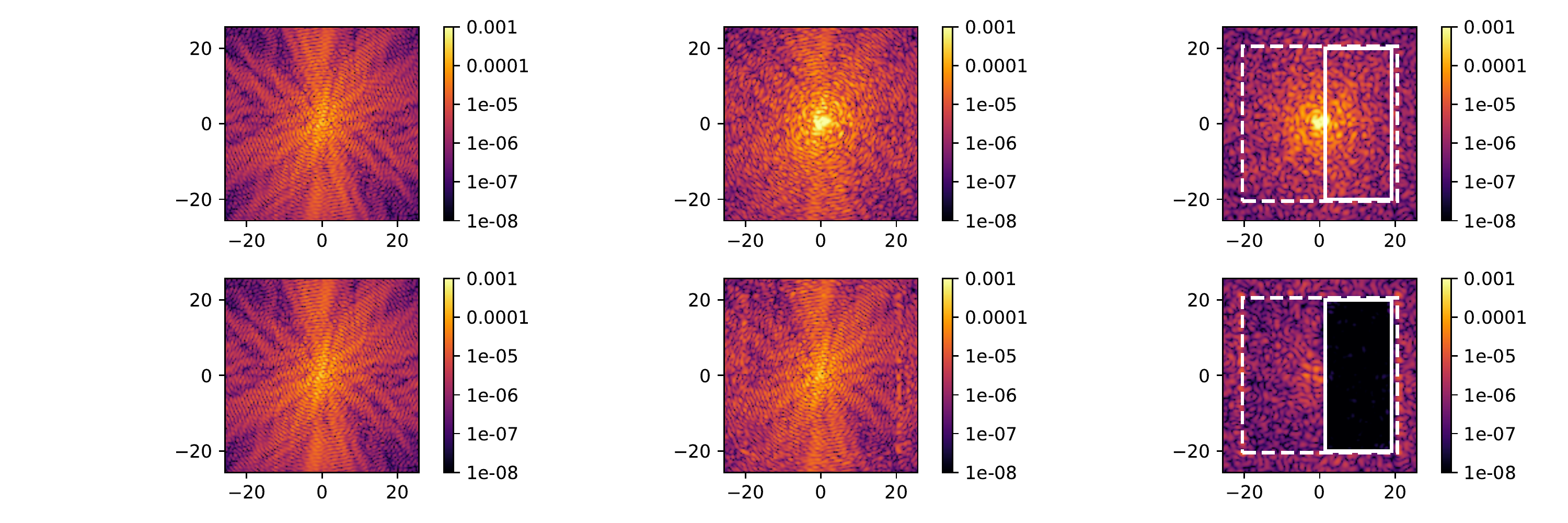}
        \caption{Focal-plane contrast before and after wavefront control for a zeroth magnitude star. The images in the top row are before wavefront control, while the images on the bottom are after wavefront control. This simulation contains both amplitude (20\,\%) and phase (75 nm ptv) aberrations. Only a one-sided dark hole can be created because there is only a single DM in the system. The control loop ran for 25 iterations, and the final dark hole contrast is well below $10^{-8}$ for channel 3. The contrast in the dark holes of channels 1 and 2 are limited by the light from the pinhole.}
        \label{fig:multi_band_control}
\end{figure*}

A one-sided dark hole was generated because both amplitude and phase errors were present. The control matrix was again generated through the WLS approach. The dark hole region in this case was a rectangular region, again with a size of $17.5 \lambda/D \times 40 \lambda / D$, with its edge at $1.5 \lambda / D$. The control region was still $40 \lambda/D \times 40 \lambda / D$. \editedn{Only a zeroth magnitude star is considered for this simulation.} The before-and-after wavefront control snapshot can be seen in Fig. \ref{fig:multi_band_control}. The reference field is now a complex pattern because it is created from random arcs that interfere with one another. This increases the total throughput of the reference field to 2 percent. This is almost 280 times larger than that of a classic SCC. However, because three channels are now used, the effective S/N gain is $\sqrt{280/3} \approx 9.7$. This particular example has a throughput that is 280 times larger; however, during the tests there were realizations with a 600 times larger throughput than that of the classic SCC. And even though the pattern is complex, it is still possible to use wavefront control to create a dark hole due to the addition of the extra channel.

\begin{figure}
        \centering
        \includegraphics[width=\columnwidth]{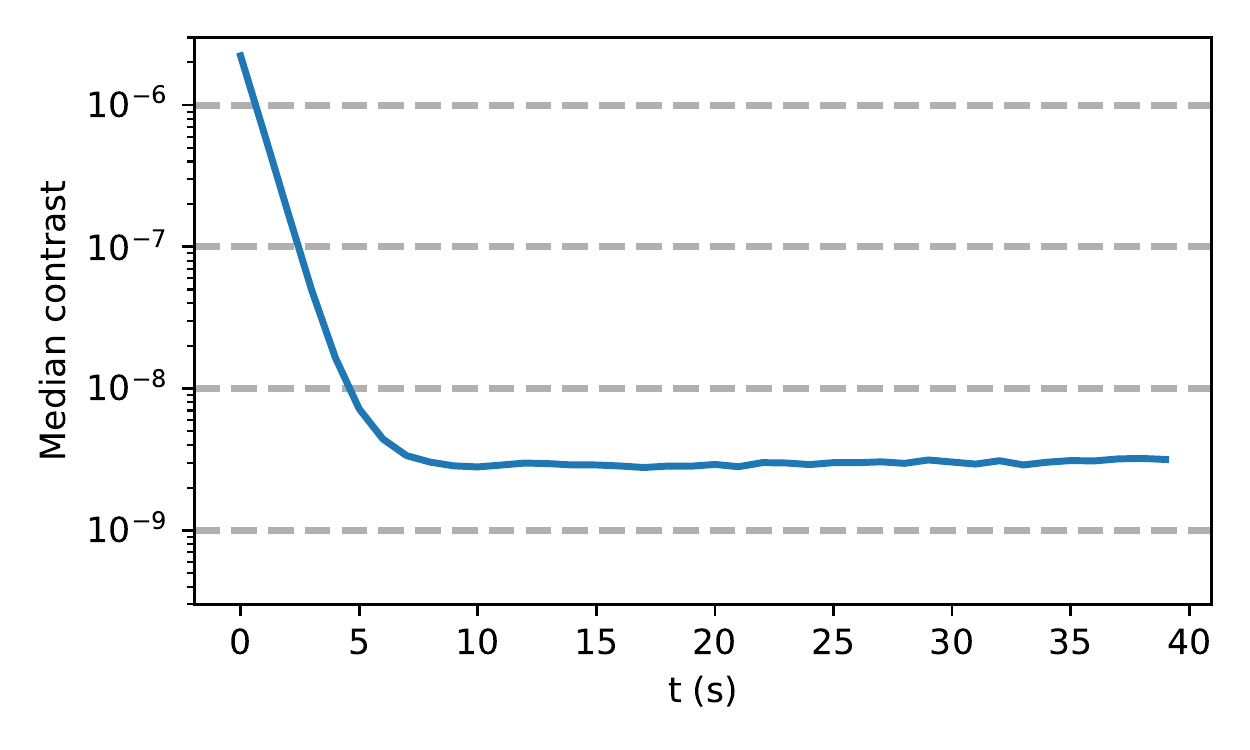}
        \caption{Median contrast in the dark hole region (1.5 $\lambda/D$ to 19.0 $\lambda/D$ in x by -20 $\lambda/D$ to 20 $\lambda/D$ in y) as a function of time. Each frame has an exposure time of 1 second. Photon noise equivalent to a zeroth magnitude star was added. The final contrast is between $2\cdot10^{-9}$ and $3\cdot10^{-9}$.}
        \label{fig:multi_band_contrast}
\end{figure}

The median contrast as a function of time is shown in Fig. \ref{fig:multi_band_control}. The final median contrast is $3\cdot10^{-9}$ (see Fig. \ref{fig:multi_band_contrast}). This is a bit higher than for the one-sided DBI case ($1\cdot10^{-9}$), even though one would expect a better contrast based on the throughput of the reference field. To create unique focal-plane probes, the amplitude modification needs to be asymmetric \citep{martinache2013afwfs, bos2019vappwfs}. The LSs in Fig. \ref{fig:multi_band_apertures} were randomly generated; this can create a partially symmetric pattern. The symmetric part will mostly add noise. This could be an explanation as to why the contrast is higher than in the pinhole case.

\section{Discussion and conclusion}
I present a new variant of the SCC, the SM-SCC. The SM-SCC modulates the pinhole and/or the LS along the spectral direction with a spectral filter. By imaging several spectral channels, either in a DBI setup or with an IFU, the fringed focal-plane image can be retrieved. In the DBI case, one of the channels will have an opaque pinhole, while the other channel's pinhole is transparent. The difference between the two OTFs reveals the sideband that carries the aberration information. The sideband can be used to sense and control the wavefront. The SM-SCC removes the limitation of the pinhole distance of the classic SCC because it is no longer necessary to resolve the sideband from the central OTF. The pinhole can be placed directly next to the aperture, which increases the throughput. The smaller pinhole distance also removes the need to oversize the optics. This makes it easier to be implemented in current HCI systems, which may have optics that cannot meet the SCC size requirements.

The \edited{properties of the SM-SCC were numerically explored in combination with} the vortex coronagraph and a clear circular aperture. From the end-to-end simulations, I found that the throughput of the SM-SCC is a factor of 32 higher than that of the SCC. This includes the loss of signal in one of the two spectral channels from the pinhole being blocked. I have shown the effects of DAs between different channels and the effect of the wavelength separation between the two channels. The SM-SCC can handle a separation of at least $\Delta \lambda / \lambda=0.1$. Differential aberrations can be accounted for in the calibration if they are static. Dynamic DAs may reduce the sensing quality, \edited{which makes it necessary to minimize them.}

The closed-loop performance was tested on a system with phase-only aberrations and a system with phase and amplitude aberrations. The SM-SCC works in both situations and reached a contrast below $1\cdot10^{-7}$ for a 7.5th magnitude star. The simulations of the system with phase and amplitude aberrations revealed that the chromatic behavior of the speckles may limit the final contrast that is reached. I derived how the chromatic behavior of speckles can be compensated for in post-processing, but this assumed speckles created by OPD errors. Amplitude errors have a different chromatic behavior, which will limit the control at the $10^{-9}$ to $10^{-8}$ level. This is well below the requirement that is necessary for ground-based instruments.

Adding more spectral channels, such as imaging with an IFU, can relax the requirement on the shape of the pinhole. With at least three channels, it is possible to reconstruct the speckle image, the fringe image, and the reference channel image. This creates the opportunity to change the shape of the reference channel LS. A three-channel IFU with a randomly generated LS was simulated to test this concept. The closed-loop performance was slightly worse than the system with only a pinhole ($3\cdot10^{-9}$ vs. $1\cdot10^{-9}$). This difference in performance is most likely caused by the random generation of the reference LS.

Realistic implementations of the SM-SCC will need to consider the effect of coherence. The light passing through the LS will see a different spectral filter than the reference field. Any difference in thickness and refractive index between the filters creates an OPD difference between the two beams. The coherence length is approximately $L_{\mathrm{coh}}=R\lambda$, with $R$ the resolving power of the observation. The reference field and the stellar speckles will no longer interfere if the light becomes incoherent. For DBI observations, $R$ is typically between 30 and 100. This limits the spectral filter thickness for the H-band DBI mode on SPHERE to 50 $\upmu$m. For the H$\alpha$ DBI mode of MagAO-X, the thickness limit is 60 $\upmu$m. Typical filters consist of a stack of several $\lambda/4$ layers with a total thickness of $N\lambda/4$, where $N$ is the number of layers. Narrowband transmission filters can be made with 10 to 20 layers \citep{furman1992basics}. The thickness is then between $2.5\lambda$ and $5\lambda$, which is significantly shorter than the coherence length of either DBI mode, and such filters are well within current manufacturing capabilities.

The SM-SCC is similar to other modulation schemes, such as the FMSCC \citep{martinez2019fmscc} and \edited{the polarization-encoded SCC \citep[PESCC;][]{bos2021polarization}}. The main difference between the three techniques is what property of light is used for the modulation. A classic SCC encodes the wavefront information in the spatial content, but this requires a pinhole far from the edge and a denser sampling of the focal plane. The FMSCC modulates the pinhole in time, which encodes the wavefront information in a time series, and the PESCC uses the polarization property of light to encode wavefront information. An advantage of the SCC, SM-SCC, and the PESCC over the FMSCC is that the electric field can be sensed instantaneously. Any speckle that changes on timescales shorter than the integration time will lower the performance. For faint objects that require long exposure time for sufficient S/N, the FMSCC will have lower performance because the speckles are not frozen. For the SM-SCC and the PESCC, this will matter less because the time-averaged speckle halo is subtracted in the difference image. The performance of the SM-SCC is quite similar to the performance of the PESCC as both have similar requirements on DAs and focal-plane sampling. The throughput gain, and therefore the S/N, is also the same. However, for the PESCC it is important that the instrumental polarization be corrected because this will create cross-talk between the measurements. Additionally, the PESCC has a similar challenge with the coherence length as the SM-SCC. The pinhole needs to have a polarizer that is thinner than the coherence length.

The simulations that were presented here are idealized, and a more realistic system will have to be explored in the future. The effects of atmospheric turbulence have not be accounted for, and more realistic coronagraphs have to be tested because they could induce chromatic behavior of speckles that cannot be calibrated. Another important aspect will be the possibility of using spectral differential imaging (SDI) and coherence differential imaging (CDI) as post-processing techniques to remove the residual speckles. This combination may work very well for the SM-SCC as SDI can remove incoherent starlight and CDI can remove the coherent part. How to combine the two techniques will be part of future work.

\begin{acknowledgements}
I would like to thank Steven Bos for the discussions about the different modulation schemes for the SCC. Support for this work was provided by NASA through the NASA Hubble Fellowship grant HST-HF2-51436.001-A awarded by the Space Telescope Science Institute, which is operated by the Association of Universities for Research in Astronomy, Incorporated, under NASA contract NAS5-26555.
\end{acknowledgements}

\bibliographystyle{aa} 
\bibliography{bibliography}

\end{document}